\documentclass[%
 reprint,
 amsmath,amssymb,
 aps,
prl,
]{revtex4-2}

\usepackage{graphicx}
\usepackage{dcolumn}
\usepackage{physics}

\makeatletter
\newcommand{\biggg}{\bBigg@{3}}
\newcommand{\vast}{\bBigg@{4}}
\newcommand{\Vast}{\bBigg@{5}}
\makeatother
\usepackage{amsmath}
\usepackage{calc}

\usepackage{color}
\usepackage{appendix}
\usepackage{ulem}

\let\oldsout\sout
\renewcommand{\sout}[1]{\textcolor{blue}{\oldsout{#1}}}

\begin{document}


\title{Generalized Bloch's Theorem for Cavity Exciton Polaron-Polaritons}

\author{Michael AD Taylor}
 \email{mike\_tay@lanl.gov}
 \affiliation{Theoretical Division, Los Alamos National Laboratory, Los Alamos, NM, 87545.}

\author{Yu Zhang}
 \email{zhy@lanl.gov}
 \affiliation{Theoretical Division, Los Alamos National Laboratory, Los Alamos, NM, 87545.}

\date{\today}

\begin{abstract}
We show that excitons coupled to cavity photons and phonons admit a generalized Bloch theorem when formulated for the conserved total crystal momentum. In minimal-coupling and Fr\"ohlich representations, the interchange of momenta between fermions and bosons breaks crystalline excitons' translational symmetry. In our symmetry-adapted frame, the Hamiltonian becomes block diagonal, without invoking approximations. The resulting formulation yields dispersions and optical responses of cavity exciton polaron-polaritons, enabling investigations that elucidate material properties in strong coupling.
\end{abstract}

\maketitle

In recent years, cavity and polariton physics have advanced dramatically, both experimentally and theoretically. We are now able to reach and probe the strong- and ultrastrong-coupling regimes for a wide variety of materials and cavity structures, ranging from organic and inorganic semiconductors to van der Waals heterostructures and molecular ensembles~\cite{Qiu2021JPCL,Li2021ARPC,Schwartz2013CPC,Thomas2016ACIE,Thomas2019S,Thomas2020N,Thomas2021,Vergauwe2019ACIE,Ebbesen2016ACR,GarciaVidal2021S,Hutchison2012ACIE,George2016PRL,Lather2019ACIE,Nagarajan2021JACS,Sau2021ACIE,Hirai2021CS,Hirai2021CL,Satapathy2021SA,Takele2021JPCC,Wiesehan2021JCP, Imperatore2021JCP,Vergauwe2019ACIE,Lather2022CS,Bayer2017NL,Yoshihara2016NP,Mueller2020N, Weight2023pccp, Bauman:2025aa}. Hybrid light--matter states created in these platforms offer powerful tools for engineering electronic structure and elementary excitations, with far-reaching implications for quantum materials~\cite{Schlawin2022APR}, quantum information science~\cite{Chng2024,Brumer1986CPL,Shapiro2012,Feist2018AP}, and micro- and optoelectronics~\cite{Mandal2023NL,Lerario2017NP,Xu2022}. 
However, as theoretical studies seek to explain increasingly more complicated physical systems in multimode and multidegree-of-freedom settings, the computational cost grows rapidly, necessitating approximations to make the calculations computationally tractable.\cite{Bernardis2018PRA,Bernardis2018PRAa,Kockum2019NRP,Mandal2023CR, Taylor2025CPR}

In condensed matter physics, Bloch's theorem is central because it drastically simplifies the complexity of calculating periodic systems by allowing each wavevector of the total Hamiltonian to be treated independently.~\cite{AshcroftMermin, Taylor2024PRB} 
However, this structure can break down upon interactions with other systems such photonic and phononic fields.~\cite{Taylor2024PRB, CohenTannoudji1997, Thirunamachandran1998,AshcroftMermin}
In minimal-coupling cavity QED the transverse vector potential does not commute with the crystal momentum unless long-wavelength or single-mode approximations are imposed, and Fr\"ohlich-type electron--phonon couplings similarly exchange momentum between the matter and lattice degrees of freedom (DOFs).
The resulting light--matter--phonon Hamiltonian is no longer block-diagonal in Bloch momentum, forcing many recent studies to resort to the site-basis under harsh approximations, thereby losing the symmetry advantages of the periodic fermionic system.~\cite{Chng2025NL,Xu2022NC, Blackham2025NL, Koshkaki2025a}

In this work, we propose a generalized Bloch's theorem for an exciton model strongly coupled to both photonic and phononic modes, giving rise to so-called exciton polaron-polaritons. This is achieved by performing a gauge-like transformation on the total system that changes the momentum frame from the exciton momentum to the polaron-polariton momentum, a generalized total crystal momentum that combines the exciton center-of-mass (CoM) motion with the photon and phonon momenta. 
The resulting framework drastically simplifies calculations of experimentally accessible observables like the dielectric function. 
By treating all DOFs on the same footing while conserving translational symmetry, this theory opens the door to quantitatively accurate and symmetry-efficient simulations of cavity-modified quantum materials.
This is particularly timely for Moir\'e superlattices and other van der Waals heterostructures,~\cite{Du2024NRP} whose very large real-space unit cells make existing polaron-polariton simulations that rely on site bases and broken translational symmetry prohibitively costly.~\cite{Chng2025NL,Xu2022NC, Blackham2025NL, Koshkaki2025a}
The theory presented in this letter, however, does not suffer from the same shortcoming, and can be applied to existing inter-layer exciton models~\cite{Slobodkin2020PRL} for Moir\'e superlattice structures, incorporating strong exciton-boson coupling without breaking the translational symmetry.

We begin with a 2D electronic Hamiltonian for an electron and a hole in an external potential as
\begin{equation} \label{eq:h_el_start}
    \hat{H}_\mathrm{el} = \sum_{j=\mathrm{e}, \mathrm{h}}\left[\frac{\hat{\bf p}_j^2}{2m_j}
    + \hat{V}(\hat{\bf x}_j) \right] 
    + \hat{U}(|\hat{\bf x}_\mathrm{h} -\hat{\bf x}_\mathrm{e}|),
\end{equation}
where $\hat{V}(\hat{\bf x}_j) = -z_j \sum_{\boldsymbol{\kappa}} w_{\boldsymbol{\kappa}} e^{i {\boldsymbol{\kappa}} \cdot  \hat{\bf x}_j}$ is the external potential operator of a 2D lattice for the $i_\mathrm{th}$ fermion with $\{ w_{\boldsymbol{\kappa}} \}$ as the set of weights in the Fourier series and ${\boldsymbol{\kappa}} \in \{ n {\bf b}_1 + n' {\bf b}_2, \{ n, n' \} \in \mathbb{Z}\}$ as the reciprocal lattice vectors for reciprocal lattice basis vectors ${\bf b}_i$. The final term in Eq.~\ref{eq:h_el_start}, $\hat{U}(|\hat{\bf x}_\mathrm{h} -\hat{\bf x}_\mathrm{e}|)=- \frac{1}{|\hat{\bf x}_\mathrm{h} -\hat{\bf x}_\mathrm{e}|}$, is the two-body electron-hole attraction term, which we approximate as the Coulomb potential for simplicity but can be extended to the Keldysh-Rytova potential for more realistic models~\cite{Shahnazaryan2025PRL,NguyenTruong2022PRB,Rytova1967MUPB,Cudazzo2011PRB,Sohier2017NL}.

Introducing CoM and relative variables,
$\hat{\bf X} = \frac{m_\mathrm{e}\hat{\bf x}_\mathrm{e} + m_\mathrm{h}\hat{\bf x}_\mathrm{h}}{M},
\hat{\bf P} = \hat{\bf p}_\mathrm{e} + \hat{\bf p}_\mathrm{h},
\hat{\bf x} = \hat{\bf x}_\mathrm{h} - \hat{\bf x}_\mathrm{e},
\hat{\bf p} = \frac{m_\mathrm{e}}{M}\hat{\bf p}_\mathrm{h}-\frac{m_\mathrm{h}}{M}\hat{\bf p}_\mathrm{e}$
with effective masses $M = m_\mathrm{e}+m_\mathrm{h}$ and $\mu = m_\mathrm{e}m_\mathrm{h}/M$, Eq.~\eqref{eq:h_el_start} becomes
\begin{equation} \label{eq:h_el-com-rel}
    \hat{H}_\mathrm{el} = \frac{\hat{\bf P}^2 }{2M} + \frac{\hat{\bf p}^2 }{2\mu} -  \frac{1}{|\hat{\bf x}|} - \sum_{\boldsymbol{\kappa}} 2i w_{\boldsymbol{\kappa}} e^{i {\boldsymbol{\kappa}} \cdot  \hat{\bf X}} \sin \Big( \frac{{\boldsymbol{\kappa}} \cdot  \hat{\bf x}}{2} \Big),
\end{equation}
(See Supplemental Material (SM) Sect.~I for details). In the absence of the final term, the CoM motion is free while the relative coordinate reduces to a 2D hydrogenic problem. The lattice couples these sectors, but only by reciprocal lattice vectors, so the exciton still admits a conventional Bloch decomposition in the CoM coordinate. 

\begin{figure} 
    \centering
    \includegraphics[width=1.0\linewidth,trim={{5 cm} {0.5 cm} {5 cm} {0.0 cm}}, clip]{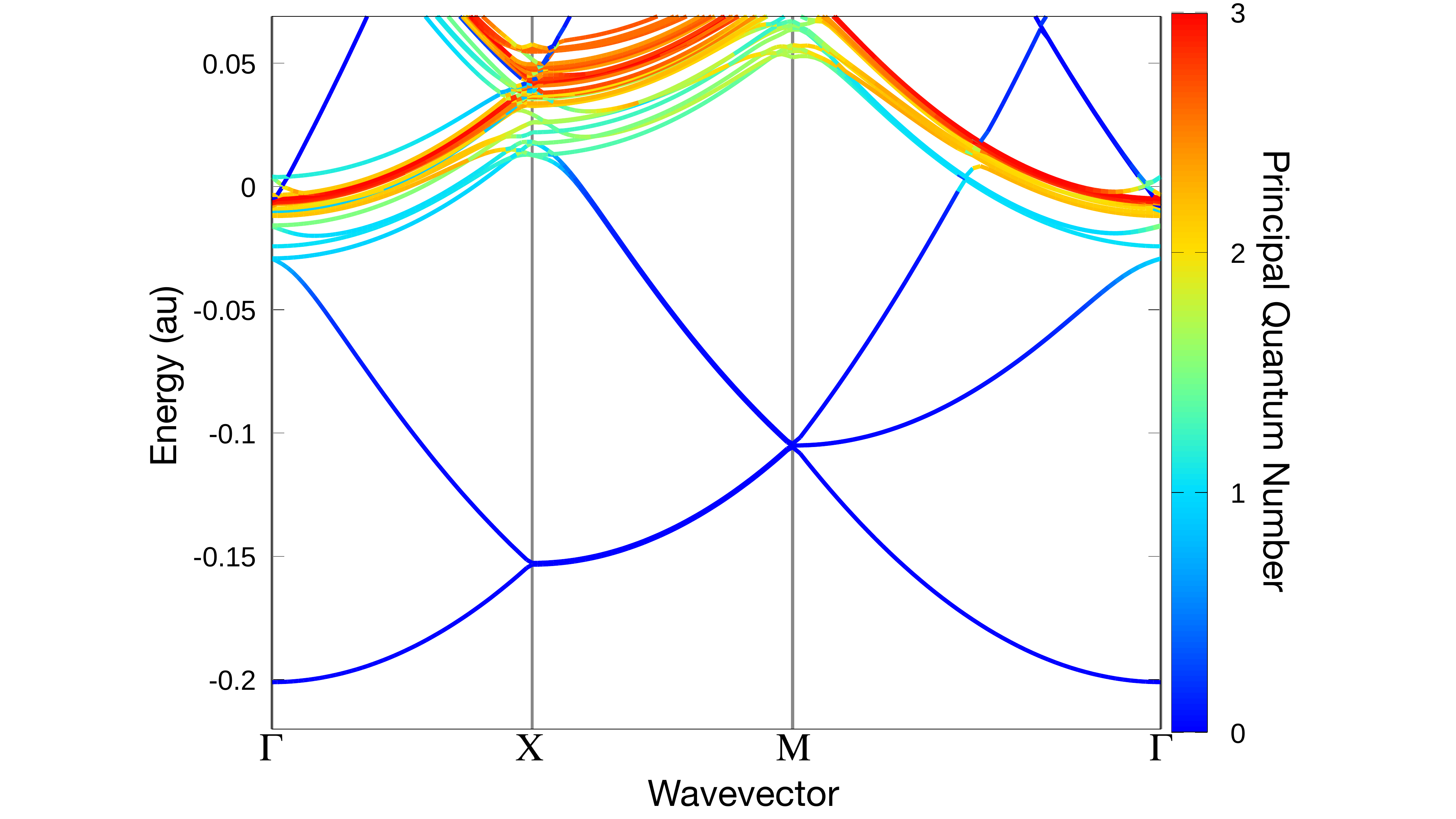}
    \caption{
    Dispersion relation of an exciton in a 2D cosine potential.
    Bands are color-coded based on the expectation value of the relative quasiparticle's principal quantum number for each state. Note how each band is duplicated and shifted for each hydrogenic state.
    Throughout this letter, the exciton parameters are $m_\mathrm{e} = m_\mathrm{h} = 0.3$au, $2 \pi / |{\bf b}_1| = 2 \pi / |{\bf b}_2| = 16$au, $w_{\boldsymbol{\kappa} \in \{\pm {\bf b}_1, \pm {\bf b}_2 \} }= 0.05$au, and all other $w_{\boldsymbol \kappa} = 0$. 
    }
    \label{fig:exciton_disp}
\end{figure}

Fig.~\ref{fig:exciton_disp} plots the exciton bandstructure for a cosine confining periodic potential. Each band is color-coded based on the average Hydrogenic principal quantum number, $\langle n \rangle$, of the state. Intuitively, we can interpret this bandstructure as a ladder of free particle (CoM quasiparticle) bands, where each band is duplicated and shifted by the energy of each Hydrogenic state (from the relative quasiparticle). The cosine potential then couples these free particle bands according to the symmetries of their corresponding Hydrogenic states, as seen in the coupling term, $-\sum_{\boldsymbol{\kappa}} 2i w_{\boldsymbol{\kappa}} e^{i {\boldsymbol{\kappa}} \cdot  \hat{\bf X}} \sin \Big( \frac{{\boldsymbol{\kappa}} \cdot  \hat{\bf x}}{2} \Big)$. Due to parity-swapping term (sine is an odd function), only opposite parity bands couple to each other, leading to the cusp at the band-edge for the lowest band. Note that this bandstructure cannot be easily modeled as a many-band nearest-neighbor tight-binding model as there is no coupling between the first two bands at the band-edge, yet there are many complicated interactions between the higher energy bands at the $\Gamma$-point.

Now we strongly couple the exciton to a multimode optical cavity through the minimal coupling Hamiltonian
\begin{align} \label{eq:h_lm-start}
    \hat{H}_\mathrm{LM} 
    =&~ \hat{H}_\mathrm{el}  + \hat{H}_\mathrm{ph} \\
    & - \sum_{j \in \mathrm{e,h}} \frac{z_j\hat{\bf p}_j \cdot \hat{\bf A}(\hat{\bf x}_j)}{m_j} + \frac{z_j^2\hat{\bf A}(\hat{\bf x}_j) \cdot \hat{\bf A}(\hat{\bf x}_j)}{2 m_j}, \nonumber
\end{align}
where $\hat{H}_\mathrm{ph} = \sum_{\bf q} \omega_{\bf q} (\hat{a}^\dagger_{\bf q} \hat{a}_{\bf q} + 1/2)$ is the photonic Hamiltonian and $\hat{\bf A}(\hat{\bf x}_j)$ is the transverse vector potential operator of the cavity field evaluated at the coordinate of the $j_\mathrm{th}$ fermion. Typically, $\hat{\bf A}(\hat{\bf x}_j)$ is decomposed into plane-wave modes as
$    \hat{\bf A}(\hat{\bf x}_j) = \sum_{\bf q} {\bf A}_{\bf q} \Big( \hat{a}_{\bf q}^\dagger e^{- i {\bf q}\cdot \hat{\bf x}_j} + \hat{a}_{\bf q} e^{ i {\bf q}\cdot \hat{\bf x}_j} \Big)$,
where ${\bf A}_{\bf q} = \sqrt{\frac{2 \pi}{\omega_{\bf q} \mathcal{V}_{\bf q}}}~{\bf e}_{\bf q}$ 
contains the vector potential amplitude and polarization, ${\bf e}_{\bf q}$, of the ${\bf q}_\mathrm{th}$ photonic mode and $\hat{a}_{\bf q}^\dagger$/$\hat{a}_{\bf q}$ are the creation/annihilation operators for the ${\bf q}_\mathrm{th}$ photonic mode with an effective mode volume $\mathcal{V}_{\bf q}$~\cite{Taylor2025a}. Unless we make the long-wavelength approximation, $[\hat{\bf p}_j, \hat{\bf A}(\hat{\bf x}_j)] \neq 0$, and since $\bf q$ is quasi-continuous for realistic cavity geometries, $\hat{\bf A}(\hat{\bf x}_j)$ breaks Bloch's theorem even for periodic systems.

To recover to the translational invariance of Bloch's theorem, we first need to transform this hybrid system to the CoM/relative frame. Upon doing so, we can rewrite Eq.~\ref{eq:h_lm-start} as
\begin{align}
    \hat{H}_\mathrm{LM} =&~ \hat{H}_\mathrm{el} + \hat{H}_\mathrm{int}^\mathrm{ph}\big(\hat{\bf p},\hat{\bf x}, \hat{\bf P}, \{ \hat{a}_{\bf q}^\dagger e^{- i {\bf q}\cdot \hat{\bf X}}\},\{ \hat{a}_{\bf q} e^{i {\bf q}\cdot \hat{\bf X}}\} \big) \\
    &+ \hat{{D}} \big(\hat{\bf x}, \{ \hat{a}_{\bf q}^\dagger e^{- i {\bf q}\cdot \hat{\bf X}}\},\{ \hat{a}_{\bf q} e^{i {\bf q}\cdot \hat{\bf X}}\} \big) + \hat{H}_\mathrm{ph} \nonumber
\end{align}
where the linear coupling term is $\hat{H}_\mathrm{int}^\mathrm{ph} \equiv \sum_{j \in \mathrm{e,h}} z_j \hat{\bf p}_j \cdot \hat{\bf A}(\hat{\bf x}_j)/m_j$, and the diamagnetic term is $\hat{{D}} \equiv \sum_{j \in \mathrm{e,h}} z_j^2 \hat{\bf A}(\hat{\bf x}_j) \cdot \hat{\bf A}(\hat{\bf x}_j) /2m_j$. Note that now $\hat{\bf X}$ only appears in the interaction terms as $\{ \hat{a}_{\bf q}^\dagger e^{- i {\bf q}\cdot \hat{\bf X}}\} /\{ \hat{a}_{\bf q} e^{i {\bf q}\cdot \hat{\bf X}}\}$.

This is now reminiscent of the single particle model from Ref.~\citenum{Taylor2024PRB}. Likewise, we introduce a new unitary operator that transforms $\hat{a}_{\bf q} e^{i {\bf q} \cdot \hat{\bf X}}\to \hat{a}_{\bf q}, \forall {\bf q}$ as 
    $\hat{U}_\mathrm{ph} \equiv \prod_{\bf q} e^{- i {\bf q}\cdot  \hat{\bf X} \hat{a}_{\bf q}^\dagger \hat{a}_{\bf q}}$.
This unitary similarly boosts the CoM momentum as
    $\hat{U}_\mathrm{ph}^\dagger \hat{\bf P} \hat{U}_\mathrm{ph} =  \hat{\bf P} - \sum_{\bf q} {\bf q} \hat{a}_{\bf q}^\dagger \hat{a}_{\bf q}$,
which can be interpreted as transforming $\hat{\bf P} \to \hat{\bf P} + \sum_{\bf q} {\bf q} \hat{a}_{\bf q}^\dagger \hat{a}_{\bf q}$, taking $\hat{\bf P}$ to now be the total electron-photon momentum. 

We can then transform our light-matter Hamiltonian by now acting $\hat{U}_\mathrm{ph}$ on it as $\hat{\tilde{H}}_\mathrm{LM}\equiv \hat{U}_\mathrm{ph}^\dagger \hat{H}_\mathrm{LM} \hat{U}_\mathrm{ph}$:
\begin{align}
     &\hat{\tilde{H}}_\mathrm{LM}= \hat{U}_\mathrm{ph}^\dagger \hat{H}_\mathrm{el}\hat{U}_\mathrm{ph} + \hat{U}_\mathrm{ph}^\dagger \hat{H}_\mathrm{int}^\mathrm{ph}\hat{U}_\mathrm{ph} + \hat{U}_\mathrm{ph}^\dagger \hat{{D}} \hat{U}_\mathrm{ph} \nonumber \\ \nonumber
     &=  \frac{\Big( \hat{\bf P} -  \sum_{\bf q} q \hat{a}_{\bf q}^\dagger \hat{a}_{\bf q} \Big)^2}{2M} + \frac{\hat{\bf p}^2 }{2\mu} - \sum_{\boldsymbol{\kappa}} 2i w_{\boldsymbol{\kappa}} e^{i {\boldsymbol{\kappa}} \cdot  \hat{\bf X}} \sin \Big( \frac{{\boldsymbol{\kappa}} \cdot  \hat{\bf x}}{2} \Big) \\ \nonumber
     & - \frac{1}{|\hat{\bf x}|}  +\sum_{\bf q} {\bf A}_{\bf q} \Bigg(\frac{2 \hat{\bf P}}{M} i \sin\Big( \frac{{\bf q}\cdot \hat{\bf x}}  {2} \Big) \big( \hat{a}^\dagger_{\bf q} - \hat{a}_{\bf q} \big) \\ \nonumber
     &- \frac{\hat{\bf p}}{\mu}\cos\Big( \frac{{\bf q} \cdot \hat{\bf x}}  {2} \Big) \big( \hat{a}^\dagger_{\bf q}  + \hat{a}_{\bf q} \big) \Bigg) \\ \nonumber
     &+ \sum_{q,q'} \frac{{\bf A}_{\bf q}\cdot{\bf A}_{\bf q'}}{2\mu} \Bigg( \cos \bigg(\frac{({\bf q} + {\bf q}') \cdot \hat{\bf x}}{2} \bigg) \Big(\hat{a}_{\bf q}\hat{a}_{\bf q'}  + \hat{a}_{\bf q}^\dagger\hat{a}_{\bf q'}^\dagger \Big) \\ 
     &+ \cos \bigg(\frac{({\bf q} - {\bf q}') \cdot\hat{\bf x}}{2} \bigg) \Big(\hat{a}_{\bf q}^\dagger\hat{a}_{\bf q'} + \hat{a}_{\bf q}\hat{a}_{\bf q'}^\dagger \Big) \Bigg) + \hat{H}_\mathrm{ph},
\end{align}
where all instances of $e^{i {\bf q}\cdot \hat{\bf X}}$ have been transformed away (See SM Sect. II). It is apparent that this form is block diagonal in the eigenbasis of $\hat{\bf P}$, restoring Bloch's theorem and allowing us to visualize the energy landscape in dispersion plots. 

This result admits a simple physical interpretation. For a bare exciton, the CoM quasiparticle satisfies Bloch's theorem even though neither the electron nor the hole do. Here, the transformation plays an analogous role: it promotes the dressed polariton to a well-defined polariton quasiparticle with momentum $\hat{\bf P}$. As all of the photon-fermion interactions are mediated through momentum exchange, by taking a step back to consider the total system, we regain the polariton momentum as a quantum number.

\begin{figure} 
    \centering
    \includegraphics[width=1.0\linewidth,trim={{5 cm} {0.5 cm} {5 cm} {0.0 cm}}, clip]{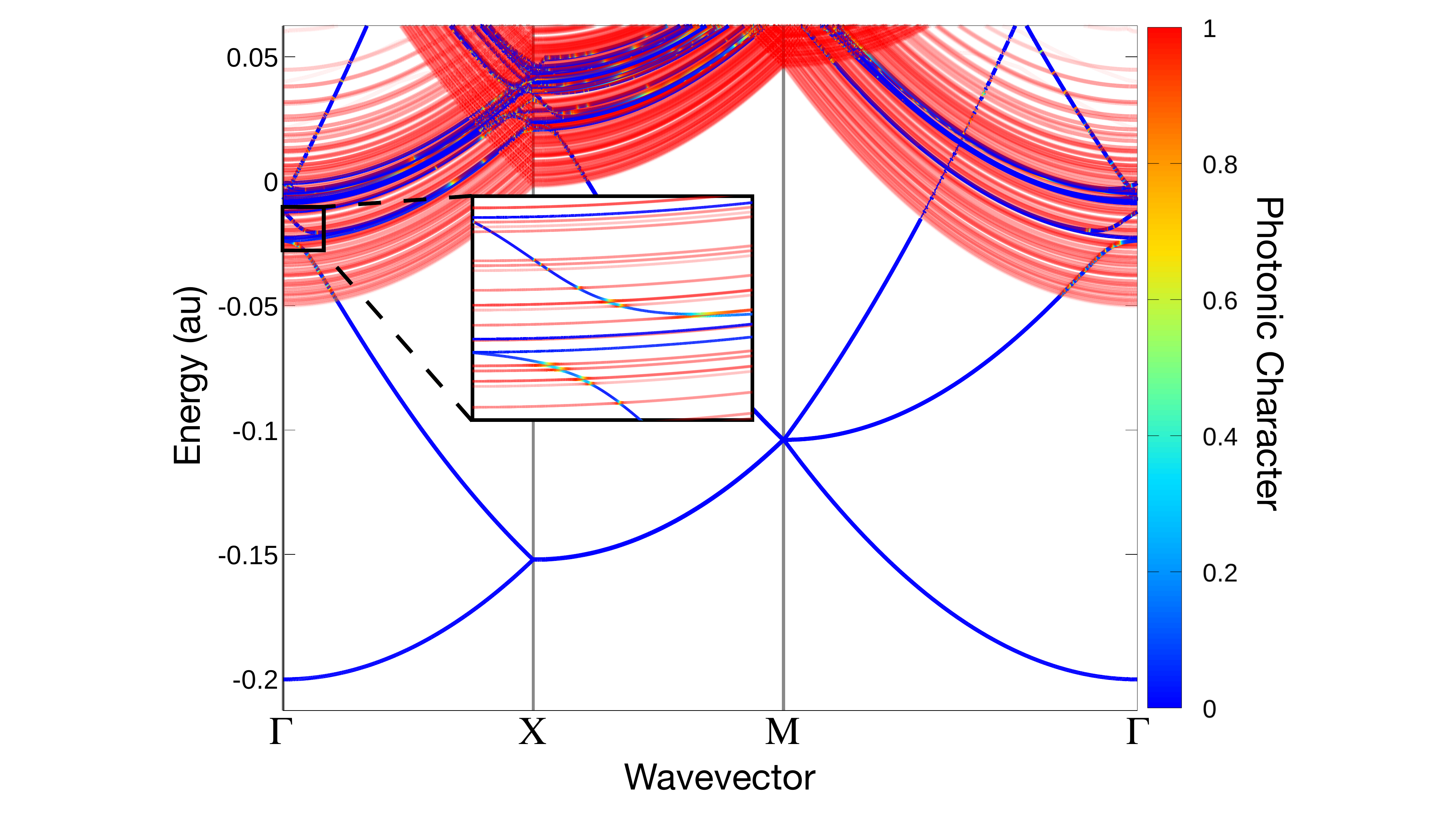}
    \caption{Dispersion relation of exciton-polariton system without any long-wavelength approximation. Bands color-coded based on photonic character. Due to the large number of modes, the transparency of the bands with a photonic character greater than 0.9 linearly increases to 95\% at pure photonic character. Photonic parameters: 2D Cartesian grid of $N = 529$ Fabry-P\'erot TE modes with $\omega_0 = 0.15$au, $\sqrt{N}A_0 = 0.003$au and a maximum $q = 0.001$au.
    }
    \label{fig:polariton_disp}
\end{figure}

In Fig.~\ref{fig:polariton_disp}, we plot the dispersion relation of the polariton momentum for 529 modes in the singly-excited subspace, on a 2D Cartesian grid in $\bf q$ for an ideal Fabry-Perot cavity. The bands are color-coded by their photonic character. Compared to the bare exciton dispersion in Fig.~\ref{fig:exciton_disp}, one can clearly observe how excitonic bands hybridize with the quasi-continuum of cavity modes, resulting in multiple avoided crossings and strongly renormalized effective masses in the polariton branches. As seen in the inset, even for bands that are seemingly unaffected, the strong coupling to many photonic modes, greatly modifies the polariton group velocity, making this a promising platform for investigating polariton transport. (See SM Sect.~III for details on matrix element calculation)

We now add in the phononic DOFs to form exciton polaron-polaritons. The total polaron-polariton Hamiltonian for this model two-electron system can be written as
\begin{equation} \label{eq:polaron_polariton_start}
    \hat{H} = \hat{\tilde{H}}_\mathrm{LM} - \sum_{j,\bf k} z_j \gamma_{{\bf k}} \Big( \hat{b}_{\bf k}^\dagger e^{-i {\bf k} \cdot \hat{\bf x}_j} + \hat{b}_{\bf k}  e^{i {\bf k} \cdot \hat{\bf x}_j} \Big) + \hat{H}_\mathrm{pn},
\end{equation}
where $\hat{b}_{\bf k}^\dagger/\hat{b}_{\bf k}$ are the creation/annihilation operators for the phonon mode with wavevector ${\bf k}$, and $\hat{H}_\mathrm{pn} = \sum_{\bf k} \omega_{\bf k} (\hat{b}_{\bf k}^\dagger\hat{b}_{\bf k} + 1/2)$ is the phonon Hamiltonian (under the harmonic mode approximation). We define the phonon-fermion coupling strength as $\gamma_{\bf k}$, such that both fermions couple identically to each $k$ phonon mode.

As with the photonic interaction terms, this Fr\"ohlich-type coupling ruins translational symmetry due to the exchange of momentum between the electronic and photonic DOFs. We then follow a similar process to try to regain Bloch's theorem. We begin by expressing the phonon-electron interaction term (second term in Eq.~\ref{eq:polaron_polariton_start}) in the CoM/relative framework as
\begin{align}
    \hat{H}_\mathrm{int}^\mathrm{pn} 
    =&~ 2 \sum_{\bf k} i \gamma_{\bf k} \sin \Big( \frac{{\bf k} \cdot \hat{\bf x}}{2} \Big) \big(\hat{b}_{\bf k}^\dagger e^{-i {\bf k} \cdot \hat{\bf X}} - \hat{b}_{\bf k} e^{i {\bf k} \cdot \hat{\bf X}}\big).
\end{align}
Once again, through this change of basis, we can now define a unitary operator to remove all of the $e^{i {\bf k} \cdot \hat{\bf X}}$ terms, restoring translational invariance for the CoM coordinate. As with the photonic DOF, the phononic unitary boost operator can be defined as
 $   \hat{U}_\mathrm{pn} = \sum_{\bf k} e^{- i {\bf k} \cdot \hat{\bf X} \hat{b}_{\bf k}^\dagger \hat{b}_{\bf k}}$,
where $\hat{U}_\mathrm{pn}^\dagger \hat{b}_{\bf k} e^{i {\bf k} \cdot \hat{\bf X}} \hat{U}_\mathrm{pn} = \hat{b}_{\bf k}$ and $\hat{U}_\mathrm{pn}^\dagger \hat{\bf P} \hat{U}_\mathrm{pn} = \hat{\bf P} - \sum_{\bf k} {\bf k}  \hat{b}_{\bf k}^\dagger \hat{b}_{\bf k}$. As such, the transformed phononic interaction Hamiltonian gets transformed to
\begin{equation}
    \hat{U}_\mathrm{pn}^\dagger \hat{H}_\mathrm{int}^\mathrm{pn} \hat{U}_\mathrm{pn} = 2i \sum_{\bf k}  \gamma_{\bf k} \sin \Big( \frac{{\bf k} \cdot \hat{\bf x}}{2} \Big) \big(\hat{b}_{\bf k}^\dagger- \hat{b}_{\bf k} \big),
\end{equation}
where all terms of $\hat{\bf X}$ are now removed.

We can now write write our total polaron-polariton Hamiltonian after these transformations as
\begin{align} \label{eq:polaron_polariton}
    \hat{\mathcal{H}} 
    =&~\frac{\Big( \hat{\bf P} -  \sum_{\bf q} {\bf q} \hat{a}_{\bf q}^\dagger \hat{a}_{\bf q} - \sum_{\bf k} {\bf k}  \hat{b}_{\bf k}^\dagger \hat{b}_{\bf k} \Big)^2}{2M} + \frac{\hat{\bf p}^2 }{2\mu} \\ \nonumber
    &- \sum_{\boldsymbol{\kappa}} 2i w_{\boldsymbol{\kappa}} \sin \Big( \frac{{\boldsymbol{\kappa}} \cdot  \hat{\bf x}}{2} \Big) +  \hat{\mathcal{H}}_\mathrm{ph} (\{ \hat{a}_{\bf q}, \hat{a}_{\bf q}^\dagger\} ) \\ \nonumber
    &+ \hat{\mathcal{H}}_\mathrm{pn} (\{ \hat{b}_{\bf k}, \hat{b}_{\bf k}^\dagger\} ) + \hat{\mathcal{H}}_\mathrm{int}^\mathrm{pn}(\hat{\bf x},\{ \hat{b}_{\bf k}, \hat{b}_{\bf k}^\dagger\} )  \\ \nonumber
    &+ \hat{\mathcal{H}}_\mathrm{int}^\mathrm{ph}(\hat{\bf P},\hat{\bf p},\hat{\bf x},\{ \hat{a}_{\bf q}, \hat{a}_{\bf q}^\dagger\} ) + \hat{\mathcal{D}} (\hat{\bf x},\{ \hat{a}_{\bf q}, \hat{a}_{\bf q}^\dagger\} ),
\end{align}
where for the sake of concision, we use the calligraphic operators to denote fully transformed operators such that $\hat{\mathcal{O}} \equiv \hat{U}_\mathrm{ph}^\dagger \hat{U}_\mathrm{pn}^\dagger \hat{O} \hat{U}_\mathrm{pn} \hat{U}_\mathrm{ph}$. It is immediately apparent that Eq.~\ref{eq:polaron_polariton} has fully restored Bloch's theorem for the CoM coordinate, $\hat{\bf X}$.

We can then parameterize $\hat{\mathcal{H}}$ by $\bf K$-points within the first Brillouin zone of the CoM particle in the plane-wave basis. We decompose a given momentum eigenstate $\ket{\bf P}$ as the planewave of a given $\bf K$-point boosted by a reciprocal lattice vector, ${\boldsymbol{\kappa}}$, such that $\ket{\bf P} = \ket{{\bf K} + {\boldsymbol{\kappa}}}$. We can then define our parameterized polaron-polariton Hamiltonian as
\begin{align} \label{eq:h_k}
    \hat{\mathcal{H}}({\bf K}) \equiv&~ \sum_{{\boldsymbol{\kappa}}, {\boldsymbol{\kappa}}'} \bra{{\bf K} + {\boldsymbol{\kappa}} + {\boldsymbol{\kappa}}'} \hat{\mathcal{H}} \ket{{\bf K} + {\boldsymbol{\kappa}}} \\ \nonumber
    =&~ \sum_{\boldsymbol{\kappa}}  \vast[ \frac{\Big( {\bf K} + {\boldsymbol{\kappa}} -  \sum_{\bf q} {\bf q} \hat{a}_{\bf q}^\dagger \hat{a}_{\bf q} - \sum_{\bf k} {\bf k}  \hat{b}_{\bf k}^\dagger \hat{b}_{\bf k} \Big)^2}{2M} + \frac{\hat{\bf p}^2 }{2\mu} \\ \nonumber
    &+  \hat{\mathcal{H}}_\mathrm{ph} + \hat{\mathcal{H}}_\mathrm{pn} + \hat{\mathcal{H}}_\mathrm{int}^\mathrm{ph}({\bf K}) + \hat{\mathcal{H}}_\mathrm{int}^\mathrm{pn} + \hat{\mathcal{D}} \vast] \hat{c}_{{\bf K} + {\boldsymbol{\kappa}}}^\dagger \hat{c}_{{\bf K} + {\boldsymbol{\kappa}}} \\ \nonumber
    &- \sum_{{\boldsymbol{\kappa}}, {\boldsymbol{\kappa}}'}  2i w_{\boldsymbol{\kappa}} e^{i {\boldsymbol{\kappa}} \cdot  \hat{\bf X}} \sin \Big( \frac{{\boldsymbol{\kappa}}' \cdot  \hat{\bf x}}{2} \Big) \hat{c}_{{\bf K} + {\boldsymbol{\kappa}} + {\boldsymbol{\kappa}}'}^\dagger \hat{c}_{{\bf K} + {\boldsymbol{\kappa}}},
\end{align}
where $\hat{c}_{{\bf K} + {\boldsymbol{\kappa}}}^\dagger / \hat{c}_{{\bf K} + {\boldsymbol{\kappa}}}$ are the creation/annihilation operators for the ${\bf K} + {\boldsymbol{\kappa}}$ plane wave of CoM particle.

Eq.~\ref{eq:h_k} is the {\it central result} of this Letter: a formally non-Bloch multimode exciton--photon--phonon problem becomes a set of independent total-crystal-momentum sectors, without assuming long-wavelength, single-mode, or site-basis approximation required by the symmetry reduction itself.
This decomposition is the key computational payoff.
This allows us to directly calculate various material properties. Namely, in this work, we will use the exact solutions of $\hat{\mathcal{H}}({\bf K})$ to calculate the dielectric function of this exciton polaron-polariton system.


\begin{figure} 
    \centering
    \includegraphics[width=1.0\linewidth,trim={{18 cm} {0.0 cm} {17.8 cm} {0.0 cm}}, clip]{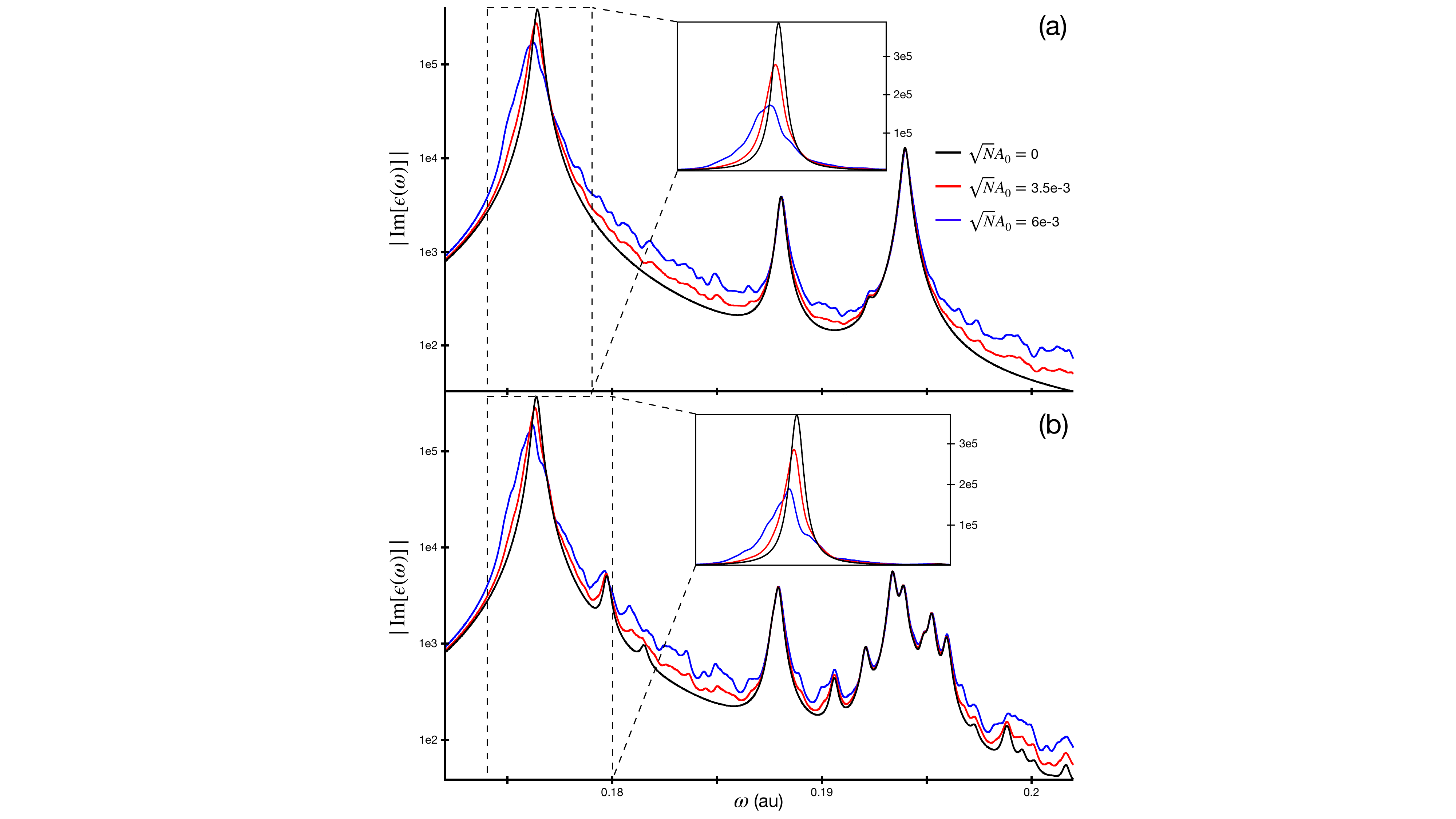}
    \caption{Imaginary component of the dielectric function for different light-matter coupling strengths under the single photon limit for (a) exciton-polariton with 2025 photonic modes and (b) exciton-polaron-polariton with 1024 photonic modes and 4 phononic Fock states. Note that imaginary component is purely negative, so its absolute value is plotted on a logarithm scale. Other exciton and photonic parameters match Fig.~\ref{fig:polariton_disp}. Phonon parameters: $\omega_{\bf k} = 500\mathrm{cm}^{-1}$, $|{\bf k}| = 1 \mathrm{nm}^{-1}$, $\gamma = 5.2\mathrm{cm}^{-1}$. Additionally, $\eta =  66\mathrm{cm}^{-1}$}
    \label{fig:pp-dielectric}
\end{figure}

To calculate the dielectric function for this hybridized system, we begin by defining the charge density function as
$\hat{\rho} (\vec{r}) = e\delta(\vec{r} - \hat{\bf x}_\mathrm{h}) - e\delta(\vec{r} - \hat{\bf x}_\mathrm{e})$,
which upon Fourier transform becomes $\hat{\rho} (\vec{q})=-2i \sin\Big(\frac{\vec{q} \cdot \hat{\bf x}}{2} \Big) e^{i \vec{q} \cdot \hat{\bf X}}$.
In this context, $\vec{q}$ will be the wavevector of the external field that will feel the dielectric function. Note that for optical frequencies, this is very small.

From linear response theory, we can then define the polarizability function of this hybrid system as
\begin{equation}
    P(\vec{q}, \omega) = \sum_{nm} \frac{| \bra{\Psi_n} \hat{\rho} (\vec{q}) \ket{\Psi_m}|^2}{\omega - (E_n - E_m) + i\eta} (f_m - f_n),
\end{equation}
where $\{ \ket{\Psi_n} \}$ are the solutions of $\hat{\mathcal{H}}({\bf K})$ with energies $E_n$ and occupations $f_n$. We also introduced a small broadening factor $\eta$ to remove singularities. Since we are doing these calculations in the Coulomb gauge, the field polarization is purely longitudinal,~\cite{Gustin2022PRA} meaning that even for this strongly-coupled system, the dielectric function calculation is the same as for a pure matter system but now using polaron-polariton states' matrix elements.

We can then directly define the dielectric function as
    $\epsilon(\vec{q}, \omega) = 1 - v(\vec{q}) P(\vec{q}, \omega)$,
where we use $v(\vec{q}) = - 4 \pi / |\vec{q}|^2$ for an unscreened Coulomb potential, since we only have a single electron-hole pair. At zero temperature and $\vec{q} \to 0$ limit, the dielectric function can be simplified as (See SM Sect.~IV):
\begin{align} \label{eq:dielectric_mat_0t}
    \epsilon_{ij}(\omega) = 1 + 4 \pi \sum_{n} \frac{\bra{\Psi_n} \hat{\bf x}_i \ketbra{\Psi_0}{\Psi_0} \hat{\bf x}_j \ket{\Psi_n}}{\omega - \Delta E_n + i\eta},
\end{align}
where $\Delta E_n = E_n - E_0$. Note that in this zero temperature case, the only state with a non-zero occupation is that of the $\Gamma$-point of the lowest exciton band. Due to the block diagonal nature of our Hamiltonian, this means that only $\hat{\mathcal{H}} (0)$ needs to be diagonalized to calculate $\epsilon_{ij}(\omega)$ at $T=0$. 

Fig~\ref{fig:pp-dielectric} plots the zero temperature imaginary component of the dielectric function of both the exciton polariton (a) and exciton polaron-polariton (b) systems for different light-matter coupling strengths.
In Fig.~\ref{fig:pp-dielectric}(a) the bare exciton curve is plotted in black. Note that the first peak contains three nearly degenerate peaks corresponding to the hybridization of the first CoM transition and the $n=0 \to n =1$ relative coordinate transitions. In each of the higher energy peaks there are many nearly degenerate higher order transitions. As we then turn on the light-matter coupling strength (red line), the higher energy peaks are largely unaffected, but the first transition couples to the photonic modes, broadening and redshifting the transition peak. As we increase the coupling strength (blue line), these effects increase and the peak shape also qualitatively changes with splitting due to light strongly coupling to one transition, hybridizing it and shifting it from the other nearly degenerate states. 

Fig.~\ref{fig:pp-dielectric}(b) highlights the additional role of strong phonon-exciton coupling.
Note that for the first exciton peak, only one transition is strongly coupled to the phonon, creating the characteristic periodic phonon sidebands. The higher energy peaks, however, show non-trivial polaron splitting and hybridization between to the many nearly-degenerate transitions. As with the phonon-free case, we see the light-matter coupling redshift and broaden the first peak. In the strong coupling case, the linewidth and shape of this peak differs from that of Fig.~\ref{fig:pp-dielectric}(a) due to one of the transitions being dressed by the phonons. Note that to lowest order in $\bf q$ and $\bf k$, the light-matter interaction scales as the matrix elements of $\hat{\bf p}$, whereas the phonon-exciton interaction scales as the matrix elements of $\hat{\bf x}$, leading to them coupling to different transitions.

In conclusion, we introduce a symmetry-informed representation for hybrid photon-exciton-phonon quantum electrodynamics (QED) Hamiltonians that restores Bloch's theorem for this hybrid system. By taking advantage of the symmetries of the interactions between these DOFs, we can create new quantum number, a so-called polaron-polariton wavevector, which turns our Hamiltonian block-diagonal in wavevector without using the long-wavelength approximation. 
The resulting block structure enables momentum-resolved dispersions and longitudinal optical response to be computed from independent ${\bf K}$ sectors, providing both conceptual clarity and a major practical simplification. More broadly, this framework supplies a natural starting point for quantitatively controlled studies of cavity-modified quantum materials, especially large-unit-cell systems such as Moir\'e heterostructures, where retaining translational symmetry is essential for tractable multimode simulations and for future extensions to nonequilibrium dynamics.

\begin{acknowledgments}
\section{Acknowledgments}
We acknowledge support from the US DOE, Office of Science, Basic Energy Sciences, Chemical Sciences, Geosciences, and Biosciences Division under Triad National Security, LLC (``Triad'') contract Grant 89233218CNA000001 (FWP: LANLECF7). This research used computational resources provided by the Institutional Computing (IC) Program and the Darwin testbed at Los Alamos National Laboratory (LANL), funded by the Computational Systems and Software Environments subprogram of LANL's Advanced Simulation and Computing program. LANL is operated by Triad National Security, LLC, for the National Nuclear Security Administration of the US Department of Energy (Contract No. 89233218CNA000001). We thank Andrei Piryatinski for fruitful discussions.
\end{acknowledgments}

\bibliography{references}


\clearpage
\begin{widetext}
\begin{center}
  {\bf  Supplemental Material for ``Restoring Bloch’s Theorem for Cavity Exciton Polaron-Polaritons"}
\end{center}
  \setcounter{equation}{0}
\renewcommand{\theequation}{S\arabic{equation}}

\section{Exciton Model}

We begin our analysis by defining a 2D electronic Hamiltonian for an electron and a hole in an external potential as
\begin{equation} \label{seq:h_el_start}
    \hat{H}_\mathrm{el} = \sum_{i=\mathrm{e},\mathrm{h}}\left[ \frac{\hat{\bf p}_i^2}{2m_i} 
    + \hat{V}(\hat{\bf x}_i)\right]
    + \hat{U}(|\hat{\bf x}_\mathrm{h} -\hat{\bf x}_\mathrm{e}|)
\end{equation}
where $\hat{V}(\hat{\bf x}_i)$ is the external potential operator for the $i_\mathrm{th}$ fermion, and $\hat{U}(|\hat{\bf x}_\mathrm{h} -\hat{\bf x}_\mathrm{e}|)=- \frac{1}{|\hat{\bf x}_\mathrm{h} -\hat{\bf x}_\mathrm{e}|}$ is the two-body electron-hole attraction term.

In this two-body case, we can exactly represent these two interacting fermions as two quasiparticles whose coordinates and momenta are defined as
\begin{align}
        \hat{\bf X} = \frac{m_\mathrm{e}\hat{\bf x}_\mathrm{e}+m_\mathrm{h}\hat{\bf x}_\mathrm{h}}{M},
        ~~~& \hat{\bf P} = \hat{\bf p}_\mathrm{e} + \hat{\bf p}_\mathrm{h} \\
        \hat{\bf x} = {\hat{\bf x}_\mathrm{h} - \hat{\bf x}_\mathrm{e}}, ~~~& \hat{\bf p} = \frac{m_\mathrm{e}}{M}\hat{\bf p}_\mathrm{h}-\frac{m_\mathrm{h}}{M}\hat{\bf p}_\mathrm{e}
\end{align}
where the center-of-mass (CoM) particle has a coordinate/momentum of $\hat{\bf X}/\hat{\bf P}$ and mass of $M = m_\mathrm{e}+m_\mathrm{h}$, and the relative particle has a coordinate/momentum of $\hat{\bf x}/\hat{\bf p}$ and mass of $\mu = m_\mathrm{e}m_\mathrm{h}/M$.For the purposes of this work, we set $m_e = m_h$.

Upon transforming to the CoM/relative frame, the kinetic energy term in Eq.~\ref{seq:h_el_start} becomes
\begin{equation}
    \frac{\hat{\bf p}_\mathrm{e}^2 + \hat{\bf p}_\mathrm{h}^2}{2m} = \frac{\hat{\bf P}^2 }{2M} + \frac{\hat{\bf p}^2 }{2\mu}
\end{equation}
where we have recovered the same form of the kinetic energy as in Eq.~\ref{seq:h_el_start} but now with two quasiparticle of different effective masses.

If we assume that the external potential is periodic, then we can express it as a Fourier series
\begin{equation}
    \hat{V}(\hat{\bf x}_j) = \sum_{\boldsymbol{\kappa}} z_j w_{\boldsymbol{\kappa}} e^{i {\boldsymbol{\kappa}} \cdot  \hat{\bf x}_j}
\end{equation}
where $\{ w_{\boldsymbol{\kappa}} \}$ is the set of weights in the Fourier series and ${\boldsymbol{\kappa}} \in \{ n {\bf b}_1 + n' {\bf b}_2, \{ n, n' \} \in \mathbb{Z}\}$ sums over the reciprocal lattice vectors for reciprocal lattice basis vectors ${\bf b}_i$. We can then write this in terms of the two quasiparticles as 
\begin{equation}
    \hat{V}(\hat{\bf x}_\mathrm{e}) + \hat{V}(\hat{\bf x}_\mathrm{h}) = -\sum_{\boldsymbol{\kappa}} 2 i w_{\boldsymbol{\kappa}} e^{i {\boldsymbol{\kappa}} \cdot  \hat{\bf X}} \sin \Big( \frac{{\boldsymbol{\kappa}} \cdot  \hat{\bf x}}{2} \Big).
\end{equation} 
By construction, this potential remains periodic for both quasiparticles in the CoM/relative frame. However, the relative coordinate's translational symmetry is broken by the two-body Coulomb term, which becomes
\begin{equation}
    \hat{U}(\hat{\bf x}) = -\frac{1}{|\hat{\bf x}|}.
\end{equation}

We can thus rewrite Eq.~\ref{seq:h_el_start} in terms of these two quasiparticles as
\begin{equation} \label{seq:h_el-com-rel}
    \hat{H}_\mathrm{el} = \frac{\hat{\bf P}^2 }{2M} + \frac{\hat{\bf p}^2 }{2\mu} -  \frac{1}{|\hat{\bf x}|} - \sum_{\boldsymbol{\kappa}} 2 i w_{\boldsymbol{\kappa}} e^{i {\boldsymbol{\kappa}} \cdot  \hat{\bf X}} \sin \Big( \frac{{\boldsymbol{\kappa}} \cdot  \hat{\bf x}}{2} \Big).
\end{equation}
Note that without the final term in Eq.~\ref{seq:h_el-com-rel}, which causes the interactions between the two quasiparticles, the CoM quasiparticle's Hamiltonian is of a free particle, and the relative quasiparticle's Hamiltonian is that of a Hydrogen atom. Due to the even parity of $\cos ( \frac{{\boldsymbol{\kappa}} \cdot  \hat{\bf x}}{2} )$, even parity relative states do not mix with odd parity states. This means that Eq.~\ref{seq:h_el-com-rel} can be diagonalized twice: once for a set of even basis functions for $\hat{\bf x}$ and once for an odd set of basis functions.

\section{Generalized Bloch's Theorem for Exciton Polariton}
If we strongly couple our exciton system to an optical cavity with many modes, we can express the minimal coupling light-matter Hamiltonian as
\begin{align} \label{seq:h_lm-start}
    \hat{H}_\mathrm{LM} =&~ \sum_{j \in \mathrm{e,h}} \frac{\Big(\hat{\bf p}_j - z_j\hat{\bf A}(\hat{\bf x}_j) \Big)^2}{2 m} + \hat{V}(\hat{\bf x}_j) + \hat{U}(|\hat{\bf x}_\mathrm{h} - \hat{\bf x}_\mathrm{e}|)  + \hat{H}_\mathrm{ph}  \nonumber\\
    =&~ \hat{H}_\mathrm{el} - \sum_{j \in \mathrm{e,h}} \frac{z_j \hat{\bf p}_j \cdot \hat{\bf A}(\hat{\bf x}_j)}{m} + \frac{z_j^2 \hat{\bf A}(\hat{\bf x}_j) \cdot \hat{\bf A}(\hat{\bf x}_j)}{2 m} + \hat{H}_\mathrm{ph} 
\end{align}
where $\hat{H}_\mathrm{ph} = \sum_{\bf q} \omega_{\bf q} (\hat{a}^\dagger_{\bf q} \hat{a}_{\bf q} + 1/2)$ is the photonic Hamiltonian and $\hat{\bf A}(\hat{\bf x}_j)$ is the transverse vector potential operator of the cavity field evaluated at the coordinate of the $j_\mathrm{th}$ fermion with charge $z_j$. Note that as stated in the main text, we assume that the electron and hole have the same mass for the purposes of this letter. Typically, $\hat{\bf A}(\hat{\bf x}_j)$ is decomposed into plane-wave modes as
\begin{equation} \label{eq:A_no_u}
    \hat{\bf A}(\hat{\bf x}_j) = \sum_{\bf q} {\bf A}_{\bf q} \Big( \hat{a}_{\bf q}^\dagger e^{- i {\bf q}\cdot \hat{\bf x}_j} + \hat{a}_{\bf q} e^{ i {\bf q}\cdot \hat{\bf x}_j} \Big)
\end{equation}
where ${\bf A}_{\bf q} = \sqrt{\frac{2 \pi}{\omega_{\bf q} \mathcal{V}_{\bf q}}}~{\bf e}_{\bf q}$ 
contains the vector potential amplitude and polarization, ${\bf e}_{\bf q}$, of the ${\bf q}_\mathrm{th}$ photonic mode and $\hat{a}_{\bf q}^\dagger$/$\hat{a}_{\bf q}$ are the creation/annihilation operators for the ${\bf q}_\mathrm{th}$ photonic mode with an effective mode volume $\mathcal{V}_{\bf q}$~\cite{Taylor2025a}. Unless we make the long-wavelength approximation, $[\hat{\bf p}_j, \hat{\bf A}(\hat{\bf x}_j)] \neq 0$, and since $\bf q$ is quasi-continuous for realistic cavity geometries, $\hat{\bf A}(\hat{\bf x}_j)$ breaks Bloch's theorem even for periodic systems.

To recover to the translational invariance of Bloch's theorem, we first need to transform this hybrid system to the CoM/relative frame. The p$\cdot$A term, $\hat{H}_\mathrm{int}^\mathrm{ph} \equiv \sum_{j = 0}^1 z_j  \hat{\bf p}_j \cdot \hat{\bf A}(\hat{\bf x}_j) / m$ from Eq.~\ref{seq:h_lm-start} can then be transformed as
\begin{align}
    \hat{H}_\mathrm{int}^\mathrm{ph} =&~ \frac{\hat{\bf P}}{M} \Big( \hat{\bf A}(\hat{\bf x}_\mathrm{e}) - \hat{\bf A}(\hat{\bf x}_\mathrm{h}) \Big) - \frac{\hat{\bf p}}{2 \mu} \Big( \hat{\bf A}(\hat{\bf x}_\mathrm{e}) + \hat{\bf A}(\hat{\bf x}_\mathrm{h}) \Big) \nonumber \\
    =&~ -\frac{2 \hat{\bf P}}{M} \bigg( \sum_{\bf q} i {\bf A}_{\bf q} \sin\Big( \frac{{\bf q}\cdot \hat{\bf x}}  {2} \Big) \big( \hat{a}^\dagger_{\bf q} e^{-i {\bf q}\cdot \hat{\bf X}} - \hat{a}_{\bf q} e^{i {\bf q}\cdot \hat{\bf X}} \big) \bigg) \nonumber \\
    &- \frac{\hat{\bf p}}{\mu} \bigg( \sum_{\bf q} {\bf A}_{\bf q} \cos\Big( \frac{{\bf q}\cdot \hat{\bf x}}  {2} \Big) \big( \hat{a}^\dagger_{\bf q} e^{-i {\bf q}\cdot \hat{\bf X}} + \hat{a}_{\bf q} e^{i {\bf q}\cdot \hat{\bf X}} \big) \bigg).
\end{align}
Note that since ${\bf A}_{\bf q} \cdot {\bf q} = 0$, we have $[{\bf A}_{\bf q} \cdot \hat{\bf p},{\bf q} \cdot \hat{\bf x}] =0 $. Now, every occurrence of $\hat{a}_{\bf q}$/$\hat{a}_{\bf q}^\dagger$ is accompanied by the same phase term, $e^{i q \hat{\bf X}}/e^{-i q \hat{\bf X}}$, respectively. 

Likewise, we can transform the diamagnetic term $\hat{{D}} \equiv \sum_{j \in \mathrm{e,h}} \hat{\bf A}(\hat{\bf x}_j) \cdot \hat{\bf A}(\hat{\bf x}_j)/2m$ to the CoM/relative frame as
\begin{align}
    \hat{{D}} =&~ \frac{1}{2m} \Bigg[\bigg(\sum_{\bf q} {\bf A}_{\bf q} \Big(e^{i{\bf q}\cdot (\hat{\bf X} + \hat{\bf x}/2)} \hat{a}_{\bf q} + e^{-i{\bf q}\cdot (\hat{\bf X} + \hat{\bf x}/2)} \hat{a}_{\bf q}^\dagger \Big) \bigg)^2 \\
    &+ \bigg(\sum_{\bf q} {\bf A}_{\bf q} \Big(e^{i{\bf q}\cdot(\hat{\bf X} - \hat{\bf x}/2)} \hat{a}_{\bf q} + e^{-i{\bf q}\cdot (\hat{\bf X} - \hat{\bf x}/2)} \hat{a}_{\bf q}^\dagger \Big) \bigg)^2 \Bigg] \nonumber \\
    =&~ \sum_{q,q'} \frac{{\bf A}_{\bf q}\cdot{\bf A}_{\bf q'}}{m} \Bigg( \cos \bigg(\frac{({\bf q} + {\bf q}') \cdot\hat{\bf x}}{2} \bigg) \Big(\hat{a}_{\bf q}\hat{a}_{\bf q'} e^{i({\bf q} +{{\bf q}'}) \cdot \hat{\bf X}} \nonumber \\
    &+ \hat{a}_{\bf q}^\dagger\hat{a}_{\bf q'}^\dagger e^{-i({\bf q} +{{\bf q}'})\cdot\hat{\bf X}} \Big) + \cos \bigg(\frac{({\bf q} - {\bf q}') \cdot \hat{\bf x}}{2} \bigg) \nonumber \\ 
    &\times  \Big(\hat{a}_{\bf q}^\dagger\hat{a}_{\bf q'} e^{i({\bf q}' - {\bf q})\cdot\hat{\bf X}} + \hat{a}_{\bf q}\hat{a}_{\bf q'}^\dagger e^{-i({\bf q} - {{\bf q}'}) \cdot \hat{\bf X}} \Big) \nonumber.
\end{align}
As with $\hat{H}_\mathrm{int}^\mathrm{ph}$, the every occurrence of $\hat{a}_{\bf q}$/$\hat{a}_{\bf q}^\dagger$ in the diamagnetic term is accompanied by the same phase term, $e^{i {\bf q}\cdot  \hat{\bf X}}/e^{-i {\bf q}\cdot  \hat{\bf X}}$, respectively. 

This is now reminiscent of the single particle model from Ref.\citenum{Taylor2024PRB}. Likewise, we introduce a new unitary operator that transforms $\hat{a}_{\bf q} e^{i q \hat{\bf X}}\to \hat{a}_{\bf q}, \forall q$ as 
\begin{equation}
    \hat{U}_\mathrm{ph} \equiv \prod_{\bf q} e^{- i {\bf q}\cdot  \hat{\bf X} \hat{a}_{\bf q}^\dagger \hat{a}_{\bf q}}.
\end{equation}
This unitary similarly boosts the CoM momentum as
\begin{equation}
    \hat{U}_\mathrm{ph}^\dagger \hat{\bf P} \hat{U}_\mathrm{ph} =  \hat{\bf P} - \sum_{\bf q} {\bf q} \hat{a}_{\bf q}^\dagger \hat{a}_{\bf q},
\end{equation}
which can be interpreted as transforming $\hat{\bf P} \to \hat{\bf P} + \sum_{\bf q} {\bf q} \hat{a}_{\bf q}^\dagger \hat{a}_{\bf q}$, taking $\hat{\bf P}$ to now be the total electron-photon momentum. 

We can then simplify our light-matter Hamiltonian by now acting $\hat{U}_\mathrm{ph}$ on it as
\begin{align}
     &\hat{U}_\mathrm{ph}^\dagger \hat{H}_\mathrm{LM} \hat{U}_\mathrm{ph} = \hat{U}_\mathrm{ph}^\dagger \hat{H}_\mathrm{el}\hat{U}_\mathrm{ph} + \hat{U}_\mathrm{ph}^\dagger \hat{H}_\mathrm{int}^\mathrm{ph}\hat{U}_\mathrm{ph} + \hat{U}_\mathrm{ph}^\dagger \hat{{D}} \hat{U}_\mathrm{ph} \nonumber \\ \nonumber
     &=  \frac{\Big( \hat{\bf P} -  \sum_{\bf q} {\bf q} \hat{a}_{\bf q}^\dagger \hat{a}_{\bf q} \Big)^2}{2M} + \frac{\hat{\bf p}^2 }{2\mu} - \sum_{\boldsymbol{\kappa}} 2 i w_{\boldsymbol{\kappa}} e^{i {\boldsymbol{\kappa}} \cdot  \hat{\bf X}} \sin \Big( \frac{{\boldsymbol{\kappa}} \cdot  \hat{\bf x}}{2} \Big) \\ \nonumber
     & +\sum_{\bf q} {\bf A}_{\bf q} \Bigg(\frac{2 \hat{\bf P}}{M} i \sin\Big( \frac{{\bf q}\cdot \hat{\bf x}}  {2} \Big) \big( \hat{a}^\dagger_{\bf q} - \hat{a}_{\bf q} \big) \\ \nonumber
     &- \frac{\hat{\bf p}}{\mu}\cos\Big( \frac{{\bf q} \cdot \hat{\bf x}}  {2} \Big) \big( \hat{a}^\dagger_{\bf q}  + \hat{a}_{\bf q} \big) \Bigg) \\ \nonumber
     &+ \sum_{q,q'} \frac{{\bf A}_{\bf q} \cdot {\bf A}_{\bf q'}}{2\mu} \Bigg( \cos \bigg(\frac{({\bf q} + {\bf q}') \cdot \hat{\bf x}}{2} \bigg) \Big(\hat{a}_{\bf q}\hat{a}_{\bf q'}  + \hat{a}_{\bf q}^\dagger\hat{a}_{\bf q'}^\dagger \Big) \\ 
     &+ \cos \bigg(\frac{({\bf q} - {\bf q}') \cdot\hat{\bf x}}{2} \bigg) \Big(\hat{a}_{\bf q}^\dagger\hat{a}_{\bf q'} + \hat{a}_{\bf q}\hat{a}_{\bf q'}^\dagger \Big) \Bigg) + \hat{H}_\mathrm{ph},
\end{align}
where all instances of $e^{i {\bf q}\cdot \hat{\bf X}}$ have been transformed away. Note that since ${\bf A}_{\bf q} \cdot {\bf q} = 0$, the $\hat{\bf P} \cdot {\bf A}_{\bf q}$ in the second line effectively remains unboosted. It is apparent that this form is block diagonal in the eigenbasis of $\hat{\bf P}$, restoring Bloch's theorem and allowing us to visualize the energy landscape in dispersion plots.  

For numerical convenience, we apply a phase rotation, $\hat{U}_{\pi/2} = \prod_{\bf q} e^{-i\frac{\pi}{2}\hat{a}_{\bf q}^\dagger\hat{a}_{\bf q} }$, to the photonic degrees of freedom such that all matrix elements when diagonalizing this are real. 
\begin{align}
     &\hat{U}_{\pi/2}^\dagger \hat{U}_\mathrm{ph}^\dagger \hat{H}_\mathrm{LM} \hat{U}_\mathrm{ph} \hat{U}_{\pi/2}= \hat{U}_\mathrm{ph}^\dagger \hat{H}_\mathrm{el}\hat{U}_\mathrm{ph} + \hat{U}_\mathrm{ph}^\dagger \hat{H}_\mathrm{int}^\mathrm{ph}\hat{U}_\mathrm{ph} + \hat{U}_\mathrm{ph}^\dagger \hat{{D}} \hat{U}_\mathrm{ph} \nonumber \\ \nonumber
     &=  \frac{\Big( \hat{\bf P} -  \sum_{\bf q} q \hat{a}_{\bf q}^\dagger \hat{a}_{\bf q} \Big)^2}{2M} + \frac{\hat{\bf p}^2 }{2\mu} - \sum_{\boldsymbol{\kappa}} 2 iw_{\boldsymbol{\kappa}} e^{i {\boldsymbol{\kappa}} \cdot  \hat{\bf X}} \sin \Big( \frac{{\boldsymbol{\kappa}} \cdot  \hat{\bf x}}{2} \Big) \\ \nonumber
     & +\sum_{\bf q} {\bf A}_{\bf q} \Bigg(\frac{2 \hat{\bf P}}{M} \sin\Big( \frac{{\bf q}\cdot \hat{\bf x}}  {2} \Big) \big( \hat{a}^\dagger_{\bf q} + \hat{a}_{\bf q} \big) \\ \nonumber
     &+i \frac{\hat{\bf p}}{\mu}\cos\Big( \frac{{\bf q} \cdot \hat{\bf x}}  {2} \Big) \big( \hat{a}^\dagger_{\bf q}  - \hat{a}_{\bf q} \big) \Bigg) \\ \nonumber
     &+ \sum_{q,q'} \frac{{\bf A}_{\bf q} \cdot {\bf A}_{\bf q'}}{2\mu} \Bigg(- \cos \bigg(\frac{({\bf q} + {\bf q}') \cdot \hat{\bf x}}{2} \bigg) \Big(\hat{a}_{\bf q}\hat{a}_{\bf q'}  + \hat{a}_{\bf q}^\dagger\hat{a}_{\bf q'}^\dagger \Big) \\ 
     &+ \cos \bigg(\frac{({\bf q} - {\bf q}') \cdot\hat{\bf x}}{2} \bigg) \Big(\hat{a}_{\bf q}^\dagger\hat{a}_{\bf q'} + \hat{a}_{\bf q}\hat{a}_{\bf q'}^\dagger \Big) \Bigg) + \hat{H}_\mathrm{ph},
\end{align}

\section{2D Hydrogen Atom}

For our relative coordinate quasiparticle, we represent its interactions with the other DOFs in total Hamiltonian in the eigenbasis of the relative Hamiltonian
\begin{align}
    \hat{H}_\mathrm{rel} = \frac{\hat{\bf p}^2}{2 \mu} - \frac{1}{|\hat{\bf x}|},
\end{align}
whose eigenfunctions in the polar coordinate system take the form
\begin{align}
    \Psi_{n,m} (r, \theta) &= C_{n,m} Y_m(\theta) R_{n,m}(r) \\
    &= C_{n,m} e^{i m \theta} (\beta_n r)^{|m|} e^{-\beta_n r /2} \mathcal{L}^{(2 |m|)}_{n-|m|} (\beta_n r) \nonumber,
\end{align}
where $n \in \mathbb{Z}^{0+}$ is the principle quantum number and $|m| \in 0,\dots,n-1$ is the angular momentum quantum number. $\mathcal{L}^{\alpha}_{n} (r)$ is the generalized Laguerre polynomial of order $n$, $\beta_n = 2\mu / (n+ 1/2)$, and the normalization constant $C_{n,m}$ takes the form
\begin{align}
    C_{n,m} = \frac{\beta_n}{\sqrt{2\pi}} \cdot \sqrt{\frac{(n-|m|)!}{(2n +1)(n+|m|)!}},
\end{align}
where the first fraction is the normalization constant for $Y_m(\theta)$ and the second one is the normalization constant for $R_{n,m}(r)$.

To build the total Hamiltonian, we need the matrix elements of the interaction terms: $(\hat{\bf p} \cdot {\bf e}_{\bf q}) \cos({\bf q} \cdot \hat{\bf x}/2)$ and $ \sin({\bf q} \cdot \hat{\bf x}/2)$. Since these integrals are not analytically accessible, we approximate these interaction terms by the leading order of their Taylor series, such that $\cos({\bf q} \cdot \hat{\bf x}/2) \approx 1 - ({\bf q} \cdot \hat{\bf x})^2/8)$ and $ \sin({\bf q} \cdot \hat{\bf x}/2) \approx ({\bf q} \cdot \hat{\bf x}/2)$. We can then write the matrix elements as
\begin{align}
    \langle  \Psi_{n',m'} | (\hat{\bf p} \cdot {\bf e}_{\bf q}) \cos({\bf q} \cdot \hat{\bf x}/2) |\Psi_{n,m} \rangle 
    \approx & -i\int d\theta \, dr \left(1 -\frac{|{\bf q}|^2 r^2}{8} \cos^2(\theta - \theta_{\bf q})\right) r  \nonumber \\
    & \times \bigg( \frac{1}{r} \sin(\theta - \theta_{{\bf e}_{\bf q}}) \Psi_{n',m'}^*(r,\theta) \frac{\partial}{\partial \theta} \Psi_{n,m}(r,\theta) \nonumber\\
    & +  \cos(\theta - \theta_{\bf q})  \Psi_{n',m'}^*(r,\theta) \frac{\partial}{\partial r} \Psi_{n,m}(r,\theta) \bigg) \nonumber \\
    \langle  \Psi_{n',m'} | \sin({\bf q} \cdot \hat{\bf x}/2) |\Psi_{n,m} \rangle  
    \approx &\int d\theta \, dr |{\bf q}| r^2 \cos(\theta - \theta_{\bf q}) \Psi_{n',m'}^*(r,\theta)  \Psi_{n,m}(r,\theta) \nonumber\\
    \langle  \Psi_{n',m'} | \cos({\bf q} \cdot \hat{\bf x}/2) |\Psi_{n,m} \rangle  
    \approx &~ \delta_{n,n'} \delta_{m,m'} - \frac{1}{8} \int d\theta \, dr |{\bf q}|^2 r^3 \cos^2(\theta - \theta_{\bf q}) \Psi_{n',m'}^*(r,\theta)  \Psi_{n,m}(r,\theta) \nonumber
\end{align}
where in polar coordinates, ${\bf q} \cdot \hat{\bf x} = |{\bf q}| \hat{r} \cos(\hat{\theta} -\theta_{\bf q})$ and 
$i\hat{\bf p} \cdot {\bf e}_{\bf q} =-\frac{1}{r} \sin(\hat{\theta} - \theta_{{\bf e}_{\bf q}}) \frac{\partial}{\partial \hat{\theta}} + \cos(\hat{\theta} - \theta_{{\bf e}_{\bf q}}) \frac{\partial}{\partial \hat{r}}$. These derivatives are evaluated as 
\begin{align}
    \frac{\partial}{\partial \theta} Y_m(\theta) =&~ im Y_m(\theta) \\
    \frac{\partial}{\partial r}  R_{n,|m|}(r) =&~ -e^{\beta_n r/2} \frac{\beta_n^{|m|+1} r^{|m|}}{2} \bigg( \Big( \frac{2|m|}{\beta_n r}  - 1 \Big)  \nonumber\\
    &\times\mathcal{L}^{(2 |m|)}_{n-|m|} (\beta_n r) - 2  \mathcal{L}^{(2 |m|+1)}_{n-|m|-1} \bigg) .
\end{align}

Using the following integral identities, we can then compute these matrix elements for arbitrary $n,n',m,m'$:
\begin{align}
    \int_0^\infty ds e^{-a s} s^{b} = & \frac{\Gamma(b +1)}{a^{b+1}}, \forall {a,b} \in \mathbb{R}^{0+}  \\
    \int_0^{2\pi} d\phi \sin(\phi - \phi') e^{i a \phi} = &
    \begin{cases}
        i \pi e^{i \phi'}, &\text{for } a = 1\\
        -i \pi e^{-i \phi'}, &\text{for } a = -1 \\
        0, &\text{otherwise}
    \end{cases}\\
    \int_0^{2\pi} d\phi \cos(\phi - \phi') e^{i a \phi} = &
    \begin{cases}
        {\pi} e^{i \phi'}, &\text{for } a = 1\\
        {\pi} e^{- i\phi'}, &\text{for } a = -1 \\
        0, &\text{otherwise}
    \end{cases} \\
    \int_0^{2\pi} d\phi \cos^2(\phi - \phi') e^{i a \phi} =& 
    \begin{cases}
        \pi , &\text{for } a = 0\\
        \frac{\pi}{2} e^{2i\phi'}, &\text{for } a = 2\\
        \frac{\pi}{2} e^{-2i\phi'}, &\text{for } a = -2 \\
        0, &\text{otherwise}
    \end{cases}\\
    \int_0^{2\pi} d\phi \sin(\phi - \phi') \cos^2(\phi-\phi'') e^{i a \phi} = 
    &\begin{cases}
        i\frac{\pi}{4} \big(2e^{i \phi'} - e^{i (2\phi'' -\phi')} \big), &\text{for } a = 1\\
        -i\frac{\pi}{4} \big(2 e^{-i \phi'} - e^{-i (2\phi'' -\phi')} \big), &\text{for } a = -1\\
        i\frac{\pi}{4} e^{i(\phi' + 2 \phi'')}, &\text{for } a = 3\\
        -i\frac{\pi}{4} e^{-i(\phi' + 2 \phi'')}, &\text{for } a = -3 \\
        0, &\text{otherwise}
    \end{cases}\\
    \int_0^{2\pi} d\phi \cos(\phi - \phi') \cos^2(\phi-\phi'') e^{i a \phi} = &\begin{cases}
        \frac{\pi}{4} \big(2e^{i \phi'} + e^{i (2\phi'' -\phi')} \big), &\text{for } a = 1\\
        \frac{\pi}{4} \big(2 e^{-i \phi'} + e^{-i (2\phi'' -\phi')} \big), &\text{for } a = -1\\
        \frac{\pi}{4} e^{i(\phi' + 2 \phi'')}, &\text{for } a = 3\\
        \frac{\pi}{4} e^{-i(\phi' + 2 \phi'')}, &\text{for } a = -3 \\
        0, &\text{otherwise}
    \end{cases},
\end{align}
which lets us solve the necessary integrals involving $R_{n,m}(r)$ as
\begin{align}
    \mathcal{O}(n,m,n',m',\ell) \equiv&~\int_0^\infty dr \, r^\ell R_{n',m'}(r) R_{n,m}(r) \\ \nonumber
    =&~ \sum_{j,j'} \beta_n^{|m| + j} \beta_{n'}^{|m'| + j'}[\mathcal{L}^{2|m|}_{n-|m|}]_j [\mathcal{L}^{2|m'|}_{n'-|m'|}]_{j'} \frac{(j + j' + |m| + |m'| + \ell)!}{\left(\frac{\beta_n + \beta_{n'}}{2}\right)^{j + j' + |m| + |m'|+ \ell + 1}} \\
    \mathcal{O}'(n,m,n',m',\ell) \equiv&~ \int_0^\infty dr \, r^\ell R_{n',m'}(r) \frac{\partial}{\partial r} R_{n,m}(r) \\ \nonumber
    =&~ {|m|} \mathcal{O}(n,m,n',m',\ell-1) - \frac{\beta_n}{2} \mathcal{O}(n,m,n',m',\ell) \\ \nonumber
    &- \sum_{j,j'} \beta^{|m|+1+j}_n \beta_{n'}^{|m'|+j'} [\mathcal{L}^{2|m|+1}_{n-|m|-1}]_j [\mathcal{L}^{2|m'|}_{n'-|m'|}]_{j'} \frac{(j + j' + |m| + |m'| + \ell)!}{\left(\frac{\beta_n + \beta_{n'}}{2}\right)^{j + j' + |m| + |m'|+ \ell +1}}
\end{align}
where we define the $[\mathcal{L}^{(a)}_{b}]_j$ such that $\mathcal{L}^{(a)}_{b}(r) = \sum_j [\mathcal{L}^{(a)}_{b}]_j\, r^j$.

Since $Y_m(\theta) = e^{im\theta}$, the non-zero matrix elements can then be written as
\begin{align}
    &i\langle  \Psi_{n',m \pm 3} | (\hat{\bf p} \cdot {\bf e}_{\bf q}) \cos({\bf q} \cdot \hat{\bf x}/2) |\Psi_{n,m} \rangle  \\ \nonumber
    &\approx C_{n,m} C_{n',m\pm3} \frac{\pi |{\bf q}|^2}{32} e^{\pm i ( \theta_\epsilon + 2 \theta_{\bf q})} \Big( \mp m \mathcal{O}(n,m,n',m \pm 3,2) \\ \nonumber
    &~~~- \mathcal{O}'(n,m,n',m\pm 3,3) \Big) 
\end{align}
\begin{align}
    &i\langle  \Psi_{n',m \pm 1} | (\hat{\bf p} \cdot {\bf e}_{\bf q}) \cos({\bf q} \cdot \hat{\bf x}/2) |\Psi_{n,m} \rangle  \\
    &\approx  C_{n,m} C_{n',m\pm1} \pi e^{\pm i \theta_\epsilon} \bigg( \pm m \mathcal{O}(n,m,n',m\pm1,0) \nonumber\\
    &~~~~+ \mathcal{O}'(n,m,n',m\pm1,1)  \nonumber \\  
    &~~~- \frac{|q|^2}{32} \Big(\mp m \big(2 - e^{\pm2i(\theta_q - \theta_\epsilon)} \big) \mathcal{O}(n,m,n',m\pm1,2) \nonumber\\
    &~~~~+ \big(2 + e^{\pm2i(\theta_q - \theta_\epsilon)} \big) \mathcal{O}'(n,m,n',m\pm1,3) \Big) \bigg) \nonumber
\end{align}
\begin{align}
    &\langle  \Psi_{n',m \pm 1} |\sin({\bf q} \cdot \hat{\bf x}/2) |\Psi_{n,m} \rangle  \\ \nonumber
    &\approx C_{n,m} C_{n',m\pm1}  \frac{\pi |{\bf q}|}{2} e^{\pm i \theta_{\bf q}} \mathcal{O}(n,m,n',m\pm1,2)
\end{align}
\begin{align}
    &\langle  \Psi_{n',m \pm 2} | \cos({\bf q} \cdot \hat{\bf x}/2) |\Psi_{n,m} \rangle  \\ \nonumber
    &\approx -C_{n,m} C_{n',m\pm2}  \frac{\pi |{\bf q}|^2}{16} e^{\mp 2 i \theta_{\bf q}} \mathcal{O}(n,m,n',m\pm2,3)
\end{align}
\begin{align}
    &\langle  \Psi_{n',m} | \cos({\bf q} \cdot \hat{\bf x}/2) |\Psi_{n,m} \rangle  \\
    &\approx  \delta_{n,n'} -C_{n,m} C_{n',m} \frac{\pi |{\bf q}|^2}{8 } \mathcal{O}(n,m,n',m,3)  \nonumber 
\end{align}

We can define the eigenenergies of this 2D Hydrogen analog as
\begin{align}
    E_n = \frac{-\mu}{2n + 1}.
\end{align}
Since the all states with a given $n$ are energetically degenerate, we decide to represent our eigenbasis as real wavefunctions, changing our angular momentum quantum number from $m \to \ell$ such that

\begin{align}
    \ket{\Phi_{n,\ell}} &= \begin{cases}
        \ket{\Psi_{n,m =0}}, &\text{for } \ell = 0\\
        \frac{\ket{\Psi_{n,m = \ell}} + \ket{\Psi_{n,m = -\ell}} }{\sqrt{2}}, &\text{for } \ell > 0\\
        \frac{\ket{\Psi_{n,m = -\ell}} - \ket{\Psi_{n,m = \ell}} }{i\sqrt{2}}, &\text{for } \ell < 0
    \end{cases}\\
    \Phi_{n,\ell}(r, \theta) &= \begin{cases}
        {\Psi_{n,m =0}}(r, \theta), &\text{for } \ell = 0\\
       \sqrt{2} C_{n,m} \cos({m \theta}) \\
        ~~~\times(\beta_n r)^{|m|} e^{-\beta_n r /2} \mathcal{L}^{(2 |m|)}_{n-|m|} (\beta_n r), &\text{for } \ell > 0\\
        \sqrt{2} C_{n,m} \sin({m \theta}) ) \\
        ~~~~\times(\beta_n r)^{|m|} e^{-\beta_n r /2} \mathcal{L}^{(2 |m|)}_{n-|m|} (\beta_n r), &\text{for } \ell < 0
    \end{cases}
\end{align}
where all even parity states have $\ell \geq 0$ and all odd parity states have $\ell < 0$. This is the basis used for our numerical simulations.

\section{Dielectric function}

To calculate the dielectric function for this hybridized system, we begin as in the main text by defining the charge density function as
\begin{equation}
    \hat{\rho} (\vec{r}) = e\delta(\vec{r} - \hat{\bf x}_\mathrm{h}) - e\delta(\vec{r} - \hat{\bf x}_\mathrm{e}),
\end{equation}
which upon Fourier transform becomes
\begin{equation}
   \hat{\rho} (\vec{q}) = e^{i \vec{q} \cdot \hat{\bf x}_\mathrm{h} } - e^{i \vec{q} \cdot \hat{\bf x}_\mathrm{e}} = -2i \sin\Big(\frac{\vec{q} \cdot \hat{\bf x}}{2} \Big) e^{i \vec{q} \cdot \hat{\bf X}}.
\end{equation}
In this context, $\vec{q}$ will be the wavevector of the external field that will feel the dielectric function. Note that for optical frequencies, this is very small.

From linear response theory, we can then define the polarizability function of this hybrid system as
\begin{equation}
    P(\vec{q}, \omega) = \sum_{nm} \frac{| \bra{\Psi_n} \hat{\rho} (\vec{q}) \ket{\Psi_m}|^2}{\omega - (E_n - E_m) + i\eta} (f_m - f_n),
\end{equation}
where $\{ \ket{\Psi_n} \}$ are the solutions of $\hat{\mathcal{H}}(K)$ with energies $E_n$ and occupations $f_n$. We also introduced a small broadening factor $\eta$ to remove singularities. Since we are doing these calculations in the Coulomb gauge, the field polarization is purely longitudinal,~\cite{Gustin2022PRA} meaning that even for this strongly-coupled system, the dielectric function calculation is the same as for a pure matter system but now using polaron-polariton states' matrix elements.

We can then directly define the dielectric function as
\begin{equation}
    \epsilon(\vec{q}, \omega) = 1 - v(\vec{q}) P(\vec{q}, \omega),
\end{equation}
where we use $v(\vec{q}) = - 4 \pi / |\vec{q}|^2$ for an unscreened Coulomb potential, since we only have a single electron-hole pair. 

For most cases, it makes sense to consider the dielectric function $\epsilon(\vec{q} \to 0, \omega) = \epsilon(\omega)$. This simplifies $\hat{\rho} (\vec{q}) \approx -i \vec{q} \cdot \hat{\bf x}$ allowing us to write $\epsilon(\omega)$ as 
\begin{align}
    \epsilon(\omega) = 1 + 4 \pi \sum_{n,m} \frac{| \bra{\Psi_n} \hat{\bf x} \ket{\Psi_m}|^2}{\omega - (E_n - E_m) + i\eta} (f_m - f_n).
\end{align}
Additionally, we can consider the dielectric matrix by decomposing $\hat{\bf x}$ into its components
\begin{align}
    \epsilon_{ij}(\omega) = 1 + 4 \pi \sum_{n,m} \frac{\bra{\Psi_n} \hat{\bf x}_i \ketbra{\Psi_m}{\Psi_m} \hat{\bf x}_j \ket{\Psi_n}}{\omega - (E_n - E_m) + i\eta} (f_m - f_n),
\end{align}
where $\hat{\bf x}_i = \hat{\bf x} \cdot \vec{v}_i, \vec{v}_i \in \{ \vec{x}, \vec{y}, \vec{z}\}$.

If we further take the temperature, $T=0$, the occupation numbers simply become $f_n = \delta_{n,0}$ simplifying our dielectric matrix to it's final form.
\begin{align}
    \epsilon_{ij}(\omega) = 1 + 4 \pi \sum_{n} \frac{\bra{\Psi_n} \hat{\bf x}_i \ketbra{\Psi_0}{\Psi_0} \hat{\bf x}_j \ket{\Psi_n}}{\omega - \Delta E_n + i\eta},
\end{align}
where $\Delta E_n = E_n - E_0$. Note that in this zero temperature case, the only state with a non-zero occupation is that of the $\Gamma$-point of the lowest exciton band. Due to the block diagonal nature of our Hamiltonian, this means that only $\hat{\mathcal{H}} (0)$ needs to be diagonalized to calculate $\epsilon_{ij}(\omega)$ at $T=0$. 
  
\end{widetext}

\end{document}